\documentclass{iaus}
\usepackage{graphicx}
\title[S 262.~~NRPs in NGC 3766] 
{Non-radial Pulsations in the Open Cluster NGC 3766}

\author[Rachael M. Roettenbacher, Ernest C. Amouzou, \& M. Virginia McSwain]   
{Rachael M. Roettenbacher$^1$, Ernest C. Amouzou$^1$,  \and \\ M. Virginia McSwain$^1$}

\affiliation{$^1$ Lehigh University, Department of Physics,  \\ 16 Memorial Drive E, Bethlehem, PA, USA
 \\ email: {\tt rmr207@lehigh.edu, eca2@lehigh.edu, mcswain@lehigh.edu} \\[\affilskip]}

\pubyear{2009}
\volume{266}  
\pagerange{xxxx}
 \date{xxxx ?? and in revised form ??}
\jname{Star Clusters}
\editors{xxxx.}
\begin{document}

\maketitle

\begin{abstract}
Non-radial pulsations (NRPs) are a proposed mechanism for the formation of decretion disks around Be stars 
and are important tools to study the internal structure of stars. NGC 3766 has an unusually large 
fraction of transient Be stars, so it is an excellent location to study the formation mechanism of Be 
star disks. High resolution spectroscopy can reveal line profile variations from NRPs, allowing 
measurements of both the degree, $l$, and azimuthal order, $m$. However, spectroscopic studies 
require large amounts of time with large telescopes to achieve the necessary high S/N and time 
domain coverage. On the other hand, multi-color photometry can be performed more easily with 
small telescopes to measure $l$ only. Here, we present representative light curves of Be stars and 
non-emitting B stars in NGC 3766 from the CTIO 0.9m telescope in an effort to study NRPs in this cluster.
\end{abstract}

\section{Introduction}

Be stars are a class of non-supergiant B-type stars with Balmer and other line emission features due to an equatorial decretion disk.  The disk is likely the result of a combination of the star's rapid rotation (near the critical limit) and non-radial pulsations (NRPs; Porter \& Rivinius 2003). 

NRPs are spherical harmonic waves traversing the surface of a star.  These pulsations can be found in multiple frequencies on the surface simultaneously (Rivinius, Baade, \& \u Stefl 2003).  There are two primary classes of NRP modes:  $g$- and $p$-modes.  $g$-modes are described by a low frequency pulsation that has gravity as its restoring force.  The dominant oscillation in this mode is transverse across the surface.  $p$-modes are dominated by high frequency, radial oscillations with a pressure restoring force (De Ridder 2001).  These modes in main-sequence, pulsating B stars are driven by the $\kappa$ mechanism (Guti\'errez-Soto et al. 2007).  

Temperature and flux gradients are established between the dimmer, cooler material on the peaks of the pulsations and the brighter, warmer material in the troughs.  The flux variations over the stellar surface are then observed as either ripples within photospheric absorption line profiles or as periodic variations in magnitude.  A large, high-resolution spectroscopic study would reveal both the degree, $l$,  and the azimuthal order, $m$, but such studies are challenging due to the need for large amounts of time on a large telescope.  Photometry, which is easily performed with data gathered by small telescopes, only measures $l$ (Rivinius, Baade, \& \u Stefl 2003). 

McSwain et al. (2008) previously showed that NGC 3766, an open cluster in Centaurus, is rich with transient Be stars.  In an effort to detect NRPs and study the formation of these transient disks, we are currently performing a long-term photometric study of the cluster.  Here we present preliminary differential light curves that reveal magnitude variations of several Be stars that are consistent with NRPs.

\section{Observations and Data Analysis}

We observed the cluster NGC 3766 using the CTIO 0.9m telescope and SITe 2048 CCD from 2008
March 19--24.  The CCD 
was used in the quad readout mode without binning.  
We used Str\"omgren $uvby$ filters and exposure 
times of $120$, $20$, $10$, and $10$ s, respectively.  No 
additional standard stars were measured.  
Sky flats were used to calibrate the $u$ and $v$ filters, and dome flats were 
used for the $b$ and $y$ filters.  
 
\begin{figure}

\hspace{-1cm}
\includegraphics[scale=0.3]{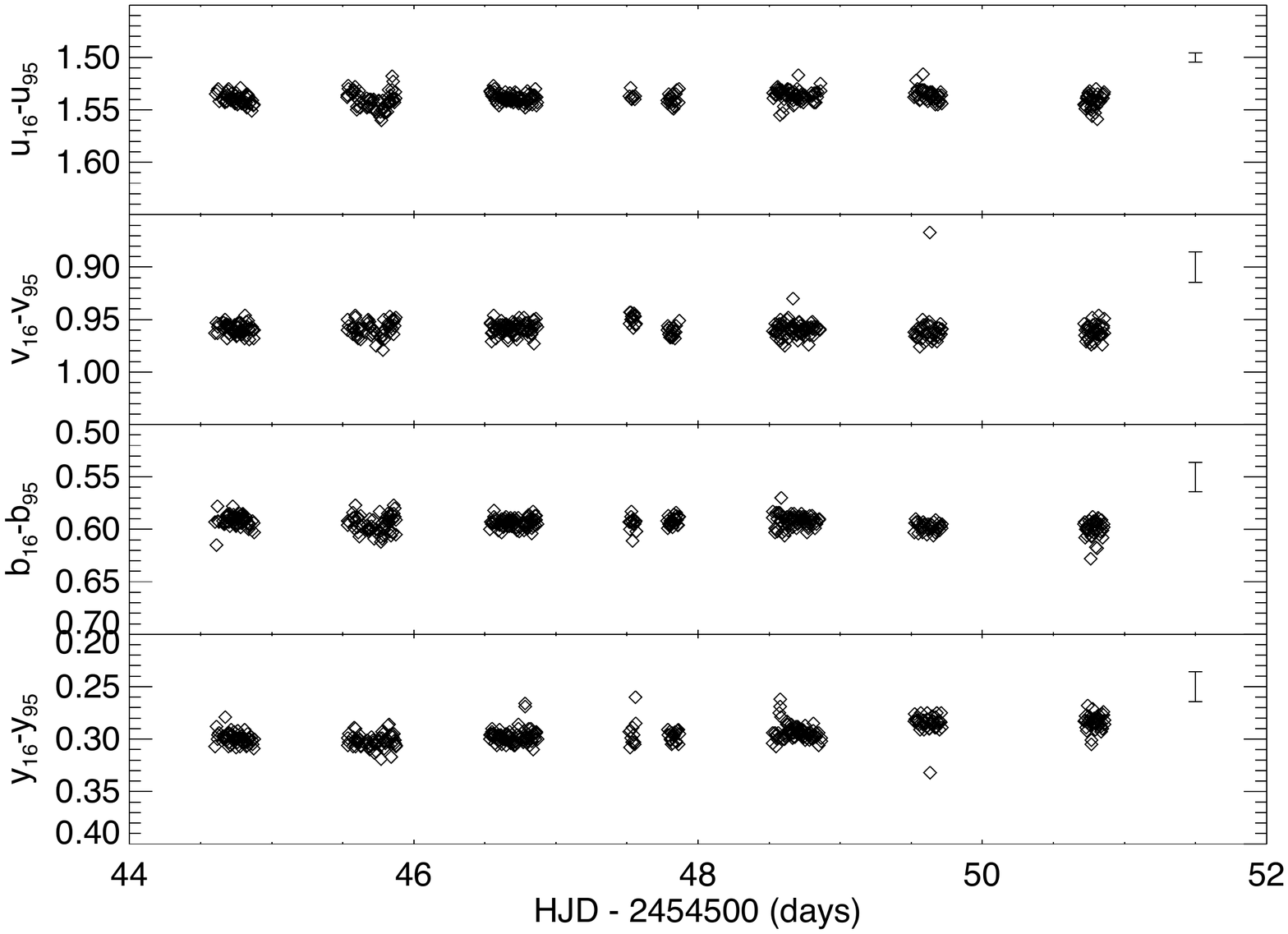}
\label{checklc}
\hspace{-1.7cm}
\includegraphics[scale=0.3]{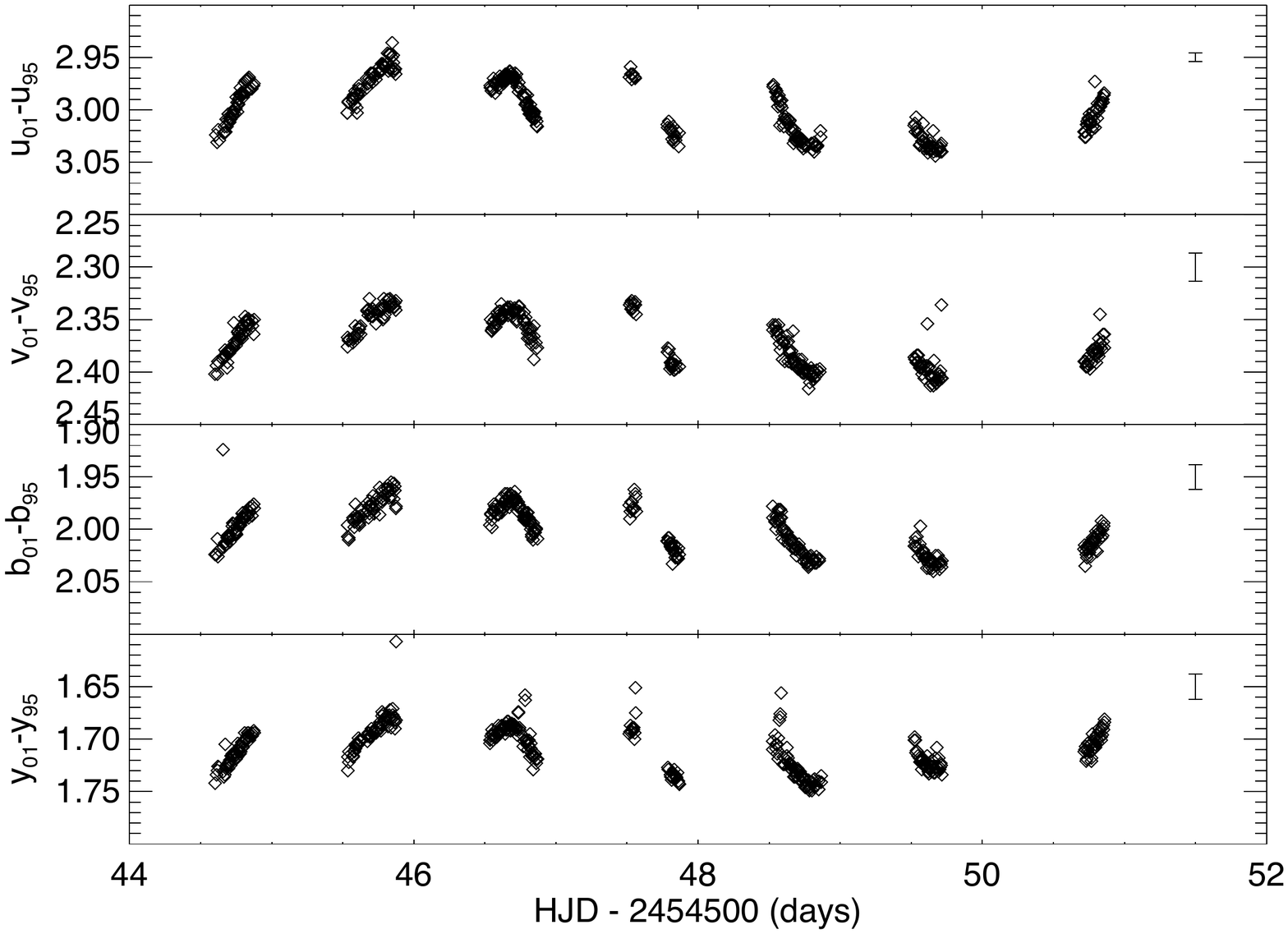}
\caption{On the left, Str\"omgren $uvby$ differential magnitudes are plotted for two check stars, Nos. 16 and 95, 
in NGC 3766.  On the right, differential magnitudes for No. 1 in the same format.  Representative error bars are shown in the upper right of each plot.  }
\label{WEBDA01lc}
\end{figure}

\begin{figure}
\hspace{-1cm}
\includegraphics[scale=0.3]{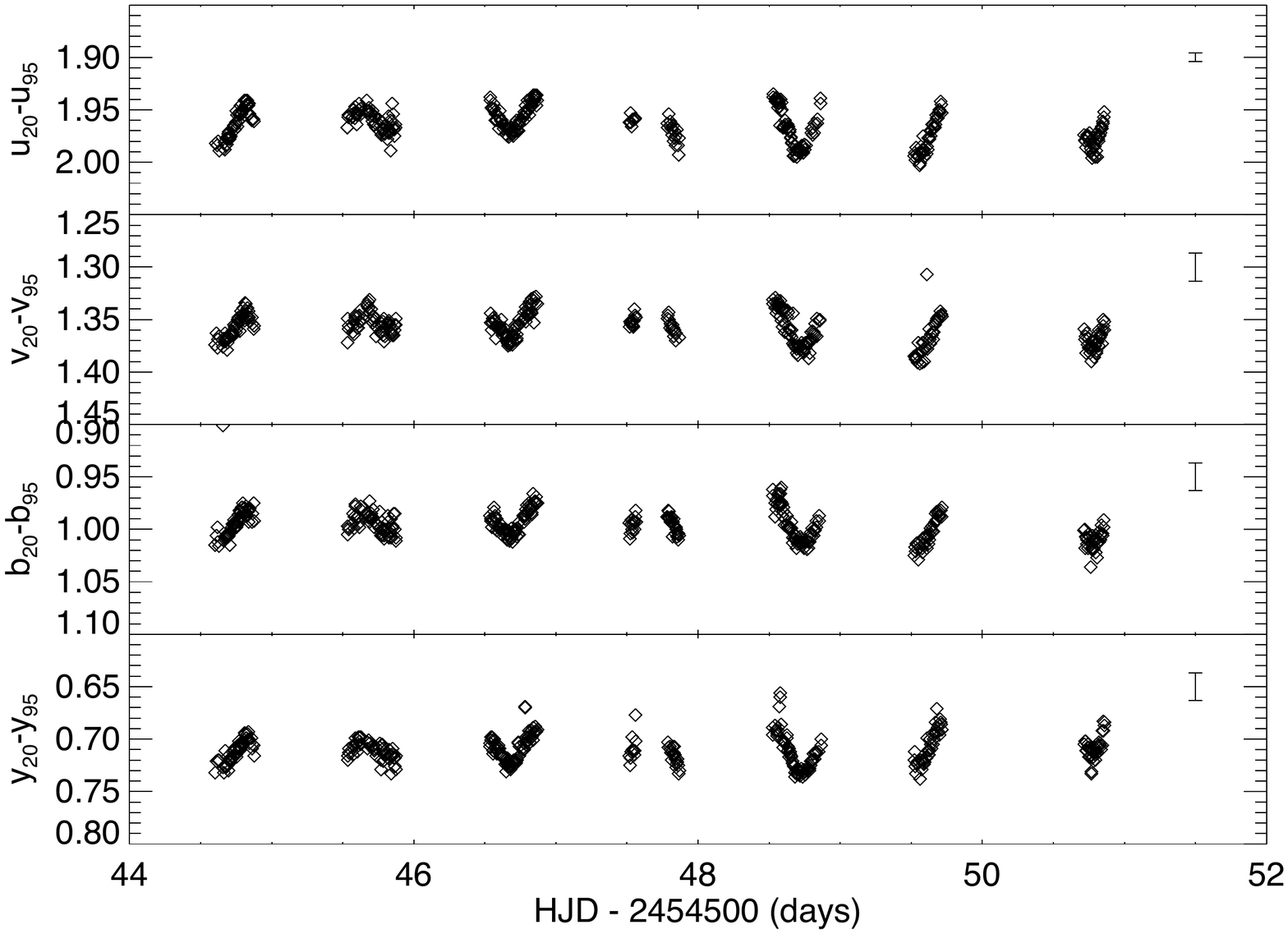}
\label{WEBDA20lc}
\hspace{-1.7cm}
\includegraphics[scale=0.3]{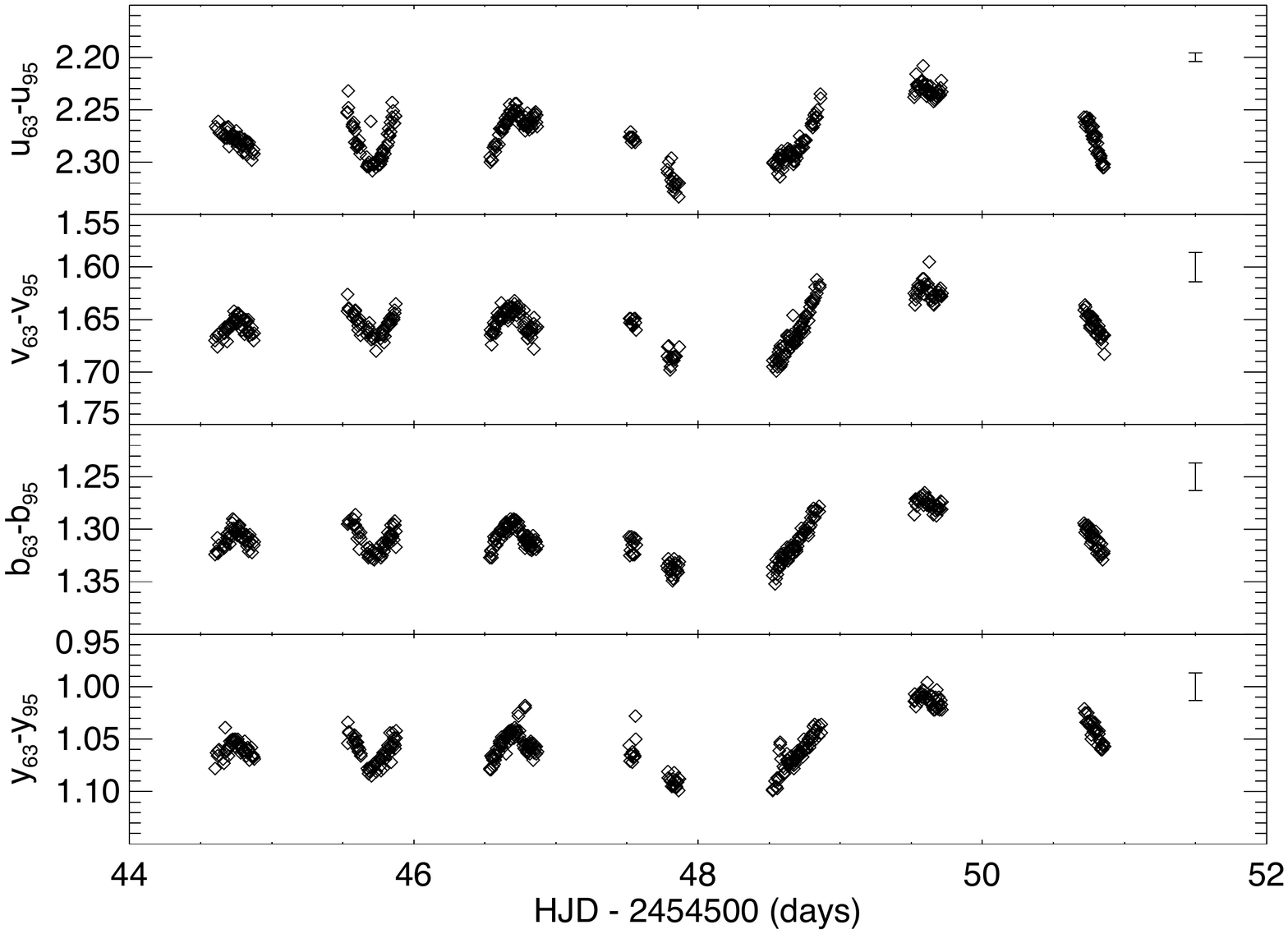}
\caption{Differential photometry for Nos. 20 and 63 in the same format as Figure \ref{WEBDA01lc}.}
\label{WEBDA63lc}
\end{figure}

The Str\"omgren $uvby$ photometric data were zero-corrected and flat field-corrected using 
 standard routines in IRAF using the $quadred$ package.  The $daofind$ and $phot$ procedures 
 were used to automatically 
identify the stars and perform aperture photometry.  
We used the numbering scheme established by Ahmed (1962) and found in the WEBDA database to identify the stars.  Differential magnitudes were determined for 
the variable Be stars identified in Balona \& Engelbrecht (1986) and 
McSwain et al.\ (2008) and four check stars, Nos. 16, 95, 111, and 147.  
The check stars have constant differential magnitudes 
within the measured errors, shown in Figure \ref{WEBDA01lc} (left).  Figures \ref{WEBDA01lc} (right) and 
\ref{WEBDA63lc} show differential light curves for the Be stars Nos. 1, 20, and 63 with 
variations in magnitude consistent with NRPs.

\section{Results and Further Work}

\begin{figure}
\centering
\includegraphics[scale=0.3]{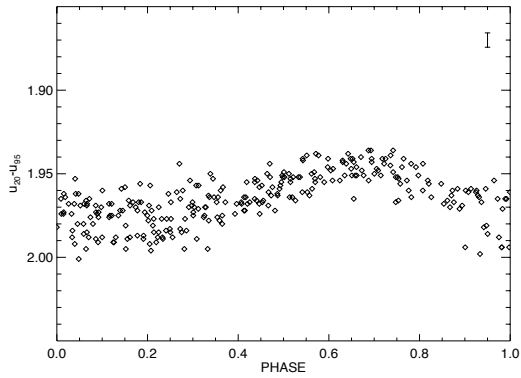}
\caption{The folded $u$-band light curve for No. 20 using a period of $6.98$ hours.}
\label{foldedlc}
\end{figure}

Nineteen of the 25 Be stars measured at this preliminary stage have magnitude variations consistent with NRPs with periods on the order of several hours.  For example, the $u$-band light curve of No. 20 was folded using a period of about $6.98$ hours, shown in Figure \ref{foldedlc}.  

\begin{figure}
\hspace{-1.5cm}
\vspace{-1.4cm}
\includegraphics[scale=0.3]{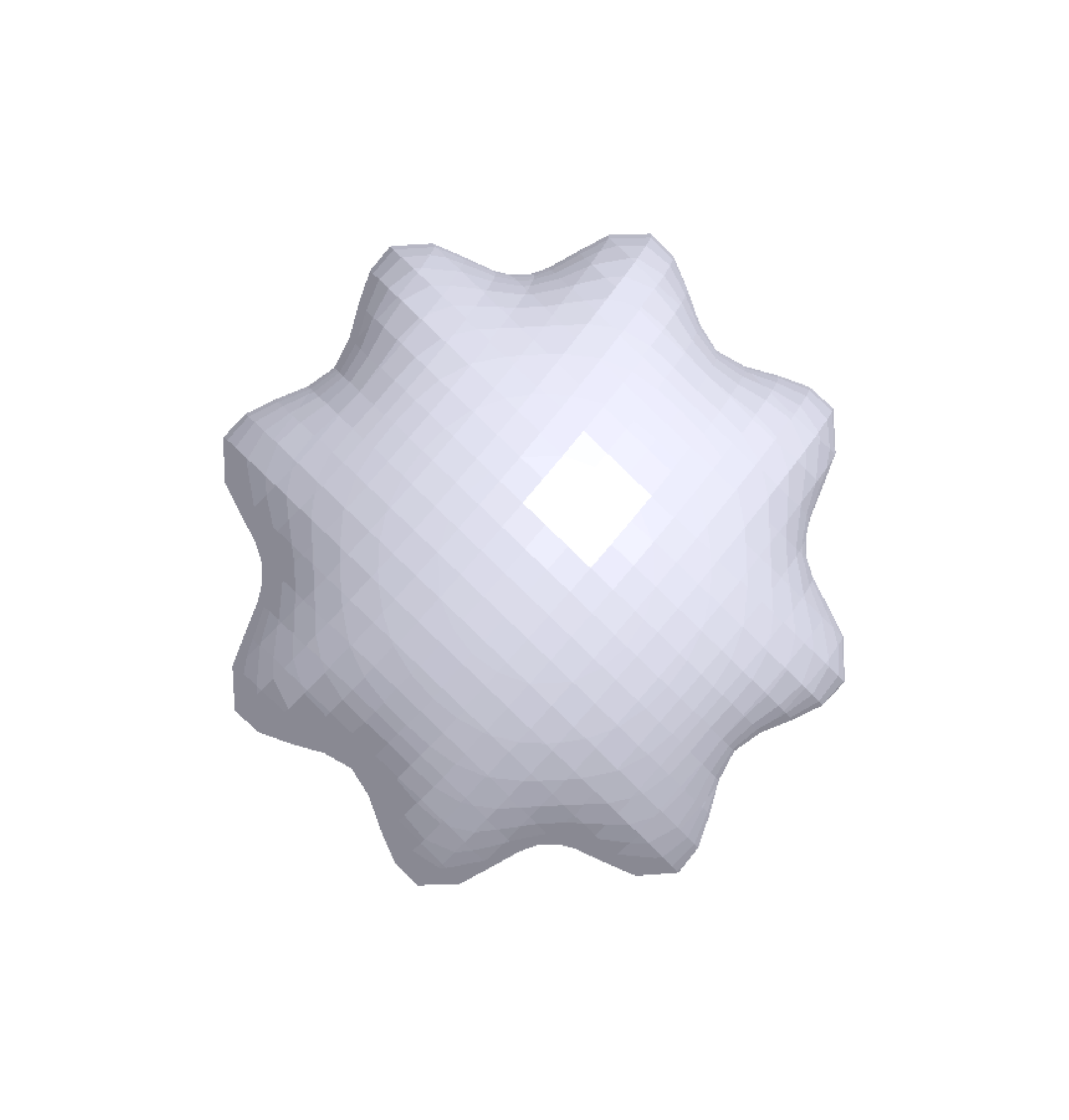}
\label{shp}
\hspace{-2cm}
\includegraphics[scale=0.3]{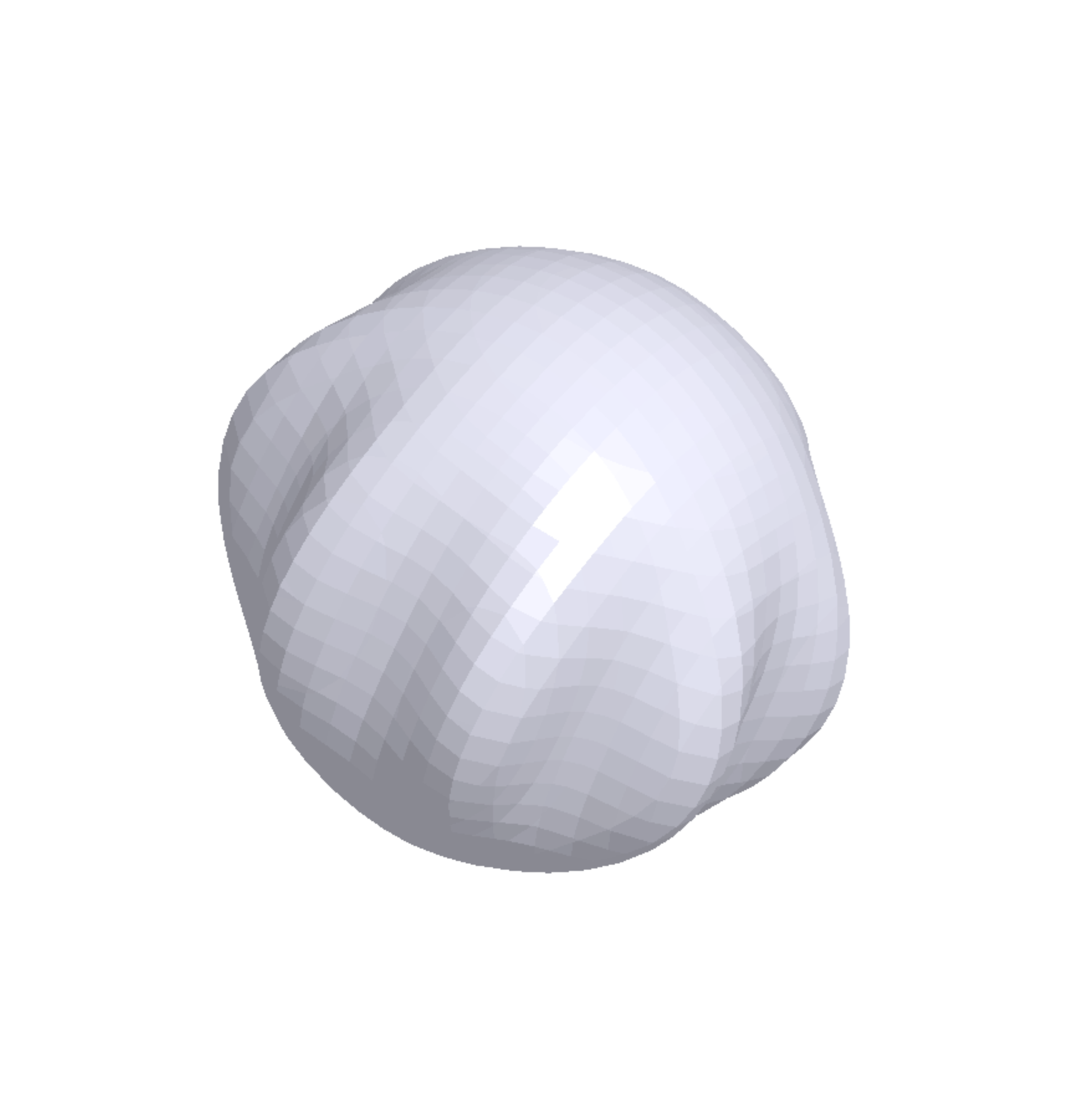}
\caption{A pole-on and an inclined model of a sphere distorted by 
non-radial pulsations, with $l  =  m  =  8$.  This is a highly exaggerated model 
of both the amplitude and high order of pulsations for display purposes.}
\label{shi}

\end{figure}

Additional data collected at the CTIO 0.9m telescope in 2008 June, 2009 February, and 2009 May on 
NGC 3766 will be analyzed in conjunction with these data to perform a period search and 
identify the various periods that are present.  In the future, we will construct a model 
for the non-radial pulsations that will allow theoretical 
light curves to be generated and compared to the observed light curves.  Assuming slowly rotating stars, the non-radial pulsations can be 
modeled by spherical harmonic perturbations of a spherical stellar surface, given by Buta \& Smith (1979) as 
\[
\frac{\delta R}{R} \thicksim A^m_{\ell} Y^m_{\ell}(\theta , \phi ) e^{\imath \sigma^m_{\ell} t} ,  
\]
where $\delta R/R$ is the fractional change in radius, $A^m_{\ell}$ is the amplitude associated with spherical harmonic $Y^m_{\ell} (\theta, \phi )$, and $\sigma ^m_{\ell}$ is the frequency of the mode.
 Pole-on and inclined examples of 
spherical harmonics (with greatly exaggerated amplitudes) are presented in Figure \ref{shi}.  The most commonly observed mode of pulsation in Be stars is $l  = | m | =  2$ (Rivinius, Baade, \& \u Stefl 2003), but 
for illustrative purposes, $l  =  m = 8$ is presented.

\section{Acknowledgements}

We gratefully acknowledge travel support from the International Astronomical Union, the American 
Astronomical Society, and NASA DPR number NNX08AV70G.  We also thank Charles Bailyn and 
the SMARTS Consortium for their help in scheduling these observations.  We are also grateful for an 
institutional grant from Lehigh University and the U.S. Department of Education GAANN Fellowship.


\begin{thebibliography}{}

\bibitem[Ahmed (1962)]{Ahmed_62}
{Ahmed, F.} 1962,
\textit{Pub. Roy. Obs. Edinburgh}, 3, 57

\bibitem[Balona \& Engelbrecht  (1986)]{Balona_Engelbrecht86}
{Balona, L. A., \& Engelbrecht, C. A.} 1986,
\textit{MNRAS}, 219, 131 

\bibitem[Buta \& Smith (1979)]{Buta_Smith79}
{Buta, R. J. \& Smith, M. A.} 1979,
\textit{ApJ}, 232, 213

\bibitem[De Ridder (2001)]{DeRidder01}
{De Ridder, J.} 2001, Ph.D. thesis,
\textit{Katholieke Univ. Leuven}

\bibitem[Gutierrez-Soto, Fabregat, Suso, Lanzara, Garrido, Hubert, \& Floquet (2007)]{Gutierrez-Soto_etal07}
{Guti\'errez-Soto, J., Fabregat, J., Suso, J., Lanzara, M., Garrido, R., Hubert, A.-M., \& Floquet, M.} 2007,
\textit{A\&A}, 476, 927

\bibitem[McSwain \etal\ (2008)]{McSwain_etal08}
{McSwain, M. V., Huang, W., Gies, D. R., Grundstrom, E. D., \& Townsend, R. H. D.} 2008, 
\textit{ApJ}, 672, 590

\bibitem[Porter \& Rivinius (2003)]{Porter_Rivinius03}
{Porter, J. M., \& Rivinius, T.} 2003,
\textit{PASP}, 115, 1153

\bibitem[Rivinus, Baade, \& \u Stefl (2003)]{Rivinius_Baade_Stefl03}
{Rivinius, T., Baade, D., \& \u Stefl, S.} 2003,
\textit{A\&A}, 411, 229




\end{thebibliography}
\end{document}